\documentclass[graybox]{svmult}

\usepackage{mathptmx}       
\usepackage{helvet}         
\usepackage{courier}        
\usepackage{makeidx}         
\usepackage{graphicx}        
\usepackage{multicol}        
\usepackage[bottom]{footmisc}

\makeindex             

\begin{document}

\title*{Super star clusters and Supernovae in interacting 
LIRGs unmasked by NIR adaptive optics}
\titlerunning{SSCs and SNe in interacting LIRGs unmasked by AO} 
\author{Petri V\"ais\"anen$^1$, Zara Randriamanakoto$^1$, Erkki Kankare$^{2,3}$,
Seppo Mattila$^2$, Stuart Ryder$^4$}
\authorrunning{V\"ais\"anen et al.} 
\institute{$^1$South African Astronomical Observatory, P.O.Box 9, Observatory 7935, Cape Town, South Africa \email{petri@saao.ac.za}\\
$^2$Tuorla Observatory, University of Turku, FI-21500 Piikki\"o, Finland\\ 
$^3$Nordic Optical Telescope, Apartado 474, E-38700 Santa Cruz de La Palma, Spain\\
$^4$Anglo-Australian Observatory, P.O.Box 296, Epping, NSW 1710, Australia
}
%
%
\maketitle


\abstract{We report on an on-going near-IR adaptive optics survey 
targeting interacting luminous IR galaxies.  High-spatial resolution NIR
data are crucial to enable interpretation of kinematic, dynamical and
star formation (SF) properties
of these very dusty objects. Whole progenitor nuclei in the interactions can 
be missed if only optical HST imaging is used. Here we specifically present
the latest results regarding core-collapse supernovae found within
the highly extincted nuclear regions of these galaxies. Direct detection and 
study of such highly obscured CCSNe
is crucial for revising the optically-derived SN rates
used for providing an independent measurement of the SF history 
of the Universe.  We also present thus-far the first NIR luminosity functions
of super star cluster (SSC) candidates.  The LFs can
then be used to constrain the formation and evolution of SSCs via constraints
based on initial mass functions and cluster disruption models.
}

\section{Introduction}
\label{sec:1}

Interacting and merging galaxies are sites of violent star formation (SF)
in extreme environments, as well as sites of the most massive known clustered 
star formation.  Many of these objects are classified as luminous or 
ultra-luminous IR galaxies (LIRGS have $\log (L_{IR}/L_{\odot}) = 11 - 12$ and 
ULIRGs $\log(L_{IR}/L_{\odot}) > 12$).
Though it is still debated exactly how much of the high-redshift SF is 
interaction triggered, and how local ULIRGs relate to their higher-$z$ cousins,
it nevertheless is clear that a significant fraction, 
if not most, of SF in the high-$z$ universe is 
happening in these extreme environments. 
It is thus important to understand the physical processes in local examples,
where detailed studies are possible, over a wide range of $L_{IR}$ and SF-rate
(SFR) output, and a range in environment and interaction stage. 

Moreover, much of the most violent SF is hidden behind dust. 
The total SF output can be estimated from far-IR
luminosities, but the spatial resolution is usually too low to see 
more details of e.g.\ the spatial distribution of SF. 
We have used NIR adaptive optics (AO) to image a sample of $<200$ Mpc
LIRGs to bridge the
gap between high-resolution optical imaging and the mid/far-IR (U)LIRG studies.
See \cite{mattila07,petri08a,petri09} for more information of the observations
and data.  We have looked for the elusive population of core-collapse SNe 
and for populations of massive super star clusters (SSCs), 
in addition to studying the galaxies
themselves.

\section{Super star cluster candidates} 
\label{sec:2}

\begin{figure}[b]
\includegraphics[scale=0.66]{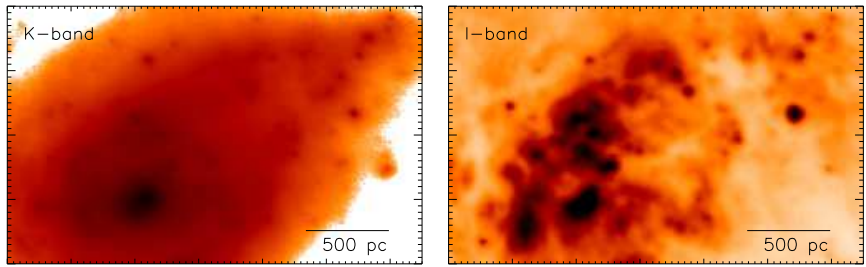}
\caption{
A region of 6''x4.5'' North-West of the 
nucleus (at bottom left) of IRAS~18293-3413 showing dozens of 
SSC candidates. 
Our VLT/NACO data is on the left and HST/ACS I-data on the right: 
SSCs are only partially detected simultaneously in the two wavelengths
because of a combination of extinction and age effects. 
}
\label{fig:1}       
\end{figure}

Most stars are born in clusters 
but most stars within galaxies do not live in clusters. 
This basic observation has led to a lively 
discussion on the physical processes involved in SSC formation and evolution, 
their disruption and survival, and
how they can be used to trace SF and 
dynamical histories of their host galaxies.
SSCs with masses even up to $10^7$ M$_{\odot}$ have 
been found in large numbers especially in interacting galaxies 
(e.g. \cite{deGrijs09} for a recent review). 
However, it still remains unclear whether the most massive clusters can only 
form in starburst environments, or are SSCs seen in those locations just
because of larger number statistics. 
SSC studies have thus far been driven by optical 
HST observations, and we now add the first significant NIR sample to be 
studied further.  
We detect from a dozen to hundreds of SSC candidates per galaxy
in our sample of LIRGs (Fig.~1). 

\subsection{The SSC NIR luminosity function}
\label{subsec:2}

The current debate over basic SSC system characteristics involves questions
such as: are the mass and luminosity functions (LF/MF) 
of SSCs universal, 
or galaxy or mass dependent, 
are the MFs/LFs described with power-laws, 
Schecter functions 
or Gaussians, 
and is SSC disruption internally driven and mass-independent or 
mass dependent and externally driven 
(see e.g. \cite{larsen09,gieles09a,fall09,elmegreen10} and 
references therein).  
There clearly is need for more data, 
preferably of new and varied targets and at different wavelengths.

\begin{figure}[t]
\includegraphics[scale=0.4]{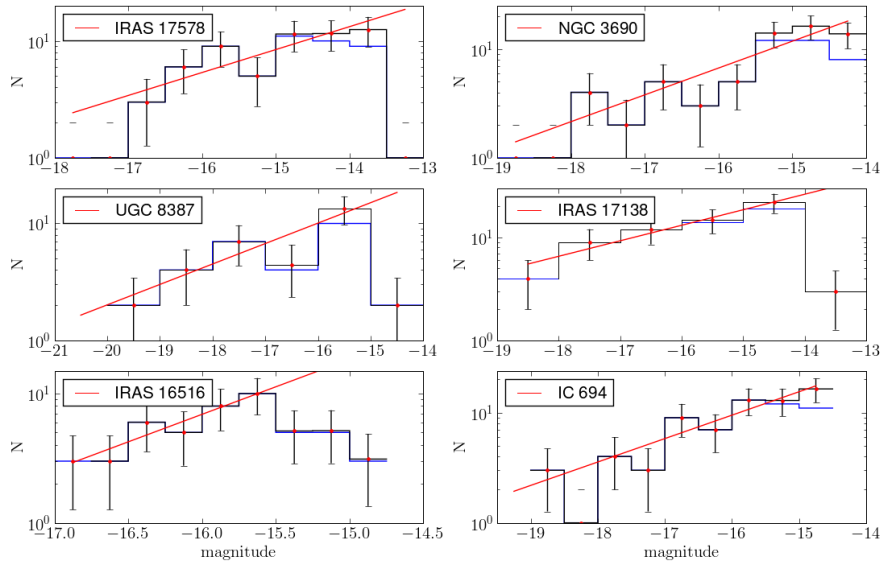}
\caption{A preliminary K-band LF of SC candidates in our sample of Gemini 
images. The black histogram and red data points show completeness corrected 
SSC number counts against K-band absolute magnitude, and the straight red lines
the weighted least-squares fit. The blue histogram is the raw SSC count.
Intriguingly, the power law LF index in these cases shows 
quite consistently $\alpha \sim -1.6$, significantly shallower than those 
seen in the optical.
}
\label{fig:2}       
\end{figure}

While waiting to complete our Gemini survey, we have already started 
putting together epoch-stacked deep images to study the LIRGs themselves.  As a 
preliminary result, Fig.~2 shows the first ever K-band luminosity function of
SSCs in external galaxies from six of our targets 
(Randriamanakoto et al., in prep.).  The LF slope alone is a useful 
diagnostic to probe SC formation/evolution since different disruption models 
result in different LFs and would indicate different initial LFs and MFs 
(e.g. \cite{gieles09b}).
Intriguingly, the fitted LF slopes are fairly consistently around 
$\alpha \sim -1.6$, i.e.\ significantly shallower than in optical studies
which typically find values of $\alpha \sim -2$.
At face value, this would indicate support for models advocating 
non-universal, mass-dependent, LFs; 
in case of a universal power law index of $\alpha\sim-2$ 
with mass-independent disruption, 
even though the observed LF is a sum of SSCs of different
ages and masses, the final LF should show the same $\alpha\sim-2$ slope.

Detecting SSCs in the K-band might be expected to result in 
different slopes given both extinction and
age effects, however -- 
in the NIR we are seeing, at least partially, a more obscured population,
and also likely a somewhat older population compared to the optically
detected SSCs.  The age difference might not necessarily be widely different,
however: our Starburst99 modelling indicates that the K-band is 
likely to pick out SSCs close to age $\sim10$ Myr, while
the optical SSCs peak only slightly earlier (excluding extinction). 
Nevertheless,  more carelul analysis of completeness
and selection effects are needed to securely reach implications from our
preliminary SSC NIR-LF results presented here.


\section{Searching for extincted core-collapse SNe}
\label{sec:3}

SF dominated (U)LIRGs are expected to hide in their central regions large 
numbers of undetected core-collapse supernovae (CCSNe), i.e.\ stars more
massive than $\sim$8~M$_{\odot}$ exploding at the end of their (short) lives. 
LIRGs with SFRs 
of tens to hundreds of M$_{\odot}$/yr can be expected to host
$\sim 0.2-1$ CCSNe/yr, a couple of orders of magnitude higher rate
than in ordinary field galaxies. Radio observations have revealed hidden SN 
factories in the nuclear regions of (U)LIRGs, e.g. 
\cite{ulvestad,miguel09}. 
Such SNe cannot be detected at optical wavelengths, even in the local universe,
because of severe ($A_V > 10$ mag) dust extinction. 
However, they {\em must} exist in the nuclear (central kpc) 
regions of such galaxies, if the high SFRs derived from their IR 
luminosities and spectra are to be believed.

Previous NIR searches \cite{mannucci03} have found an
SN-rate lower by factors of several than
expected. 
We have argued \cite{mattila04} that this rate misses the 
SNe exploding close to the centres of LIRGs in very extinguished regions,
and have an on-going program to search for these cases making use of
AO-assisted NIR imaging.


SN2004ip was detected in a pilot study using VLT/NACO \cite{mattila07}. 
It was at a projected distance of 1.4'', or 500 pc, from the nucleus of 
IRAS~18293-3413, among the closest SNe detected (in IR) to a LIRG nucleus.
Our subsequent radio observations confirmed its CCSN nature \cite{miguel07}.
Our current Gemini/ALTAIR/NIRI search has so far produced 3 new SNe. 
The first one was SN2008cs \cite{kankare08}, 
located at 4.2", or 1.5kpc, projected distance 
from the nucleus or IRAS~17138-1017. 
Follow-up observations in both radio and NIR bands were 
again consistent with a core-collapse event, and showed the SN to suffer
from a very high host galaxy extinction of $A_V \approx 16$ mag, the highest 
definite measurement yet for any SN.  In addition, a ``historical'' 
SN2004iq 
was detected in the same galaxy from HST images, and is located 660 pc from 
the nucleus. 
Earlier this year, three more SNe were detected in our Gemini programme, though
two of them were actually first seen in non-AO optical and NIR searches 
\cite{newton10,mattila10,ryder10} thanks to their relatively large 
galactocentric distances.  The third one \cite{kankare10} in IC~883, however, 
turned out to be the closest SN to a LIRG nucleus yet discovered, at just
180 pc (or 0.4'') projected distance from the core (Fig.~3).

The intrinsic number of CCSNe based on the FIR-luminosities for our 
sample of 7 galaxies over the Gemini program is calculated to be about 10 SNe.
Though the number statistics are arguably still low, and though we have not 
finished the survey, with 5 SN detections in the sample it already appears
clear that AO-assisted NIR observations provide an excellent window to detect
SNe in the obscured nuclear regions of nearby LIRGs.

\begin{figure}[t]
\includegraphics[scale=0.30]{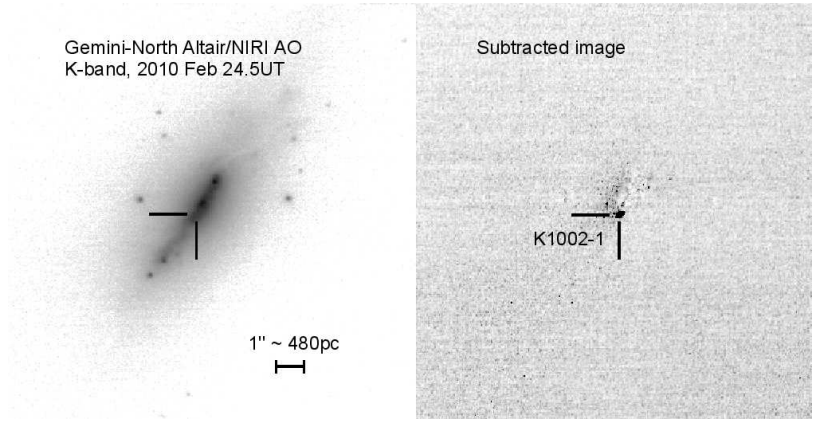}
\caption{The central regions of IC~883 observed with Gemini. The location
of the recent SN candidate PSN K1002-1 is marked, though it is  
visible only in the subtracted image from two different epochs in the 
right panel.  The projected distance to the core of the galaxy is
a mere 180 pc.
}
\label{fig:3}       
\end{figure}

\section{SF and dynamics in interacting galaxies}
\label{sec:4}

We have followed-up in detail two of our early high-quality deep NACO AO images.
In addition to combining the AO-data with archive optical HST imaging we
obtained spectroscopy using SALT and AAT. Both cases revealed surprises.
We found an unexpected third component in the supposed pair-interaction
of IRAS~19115-2124, dubbed the Bird \cite{petri08a}.  
Moreover, this third, least massive and
irregular component dominates the current star-formation output of the whole
system. This is in contrast to the widely held picture that tidal interactions
are expected to drive large quantities of gas into the {\em central regions} 
of the interaction resulting in starbursts.  While there is strong SF
in the central regions as well, it clearly is the smaller ``extra'' component
that elevates the Bird into almost a ULIRG-class object. 
We are in the process of studying this system further with integral field
spectroscopy and mid-IR imaging to piece together what has happened in this
intriguing system.

Follow-up of IRAS18293-3134, on the other hand, showed evidence of an extremely
rare leading-arm spiral, i.e.\ a galaxy where the spiral arms open up in 
the same direction as the disk is rotating \cite{petri08b}.  
Simulations have shown that some retrograde encounters
should produce these kind of galaxies \cite{thomasson}, though they have not
been studied much and only a couple of candidates exist in the literature
\cite{byrd}. 
Their very existense, however, would have implications
for e.g.\ the dark matter halo mass of spirals. 

Both of these systems highlight the fact that NIR-AO imaging is crucial in 
interpreting the dynamical state of dusty LIRGs. Relying only on optical data,
even HST-data in the I-band, will often not reveal all the components of
the interactions and will not resolve the locations of major stellar mass
distribution.  
Without the NIR AO-data it would have been virtually 
impossible to disentangle the true velocity dispersions of the nuclei from 
other kinematic components such as tidal tails and gas outflows.  
Similarly, the leading arms of IRAS~18293-3134
would not have been possible to detect without NIR AO-data, raising the
tantalizing possibility that more such cases might be hidden in the chaotic
interactions of dusty LIRGs.

\section{Summary}

We have presented results of our on-going survey to unmask CCSNe in the 
highly extincted nuclear regions of interacting luminous IR galaxies, and
presented, for the first time, luminosity functions of SSCs found in these
galaxies in the NIR.

%

%

{}


\begin{thebibliography}{}



\bibitem{byrd} Byrd G.G., Freeman T. \& Howard S., 1993, AJ, 105, 447 

\bibitem{deGrijs09} de Grijs R., 2010, RSPTA, 368, 693 (arXiv:0911.0778) 

\bibitem{elmegreen10} Elmegreen B.G. \& Hunter D.A., 2010, ApJ, 712, 604


\bibitem{fall09} Fall S.M., et al. 2009, ApJ, 704, 453 

\bibitem{gieles09a} Gieles M., 2009, MNRAS,394,2113 

\bibitem{gieles09b} Gieles M., 2010, in: B. Smith et al. (eds), Galaxy Wars, ASPC, 423, 123 (arXiv:0908.2974)


\bibitem{kankare08} Kankare E., et al., 2008, ApJ, 689, L97 

\bibitem{kankare10} Kankare E., et al.  2010, CBET, 2213 

\bibitem{larsen09} Larsen S.S. 2009, A\&A, 494, 539 


\bibitem{mannucci03} Mannucci F., et al. 2003, A\&A, 401, 519 


\bibitem{mattila04} Mattila S., Meikle W.P.S., Greimel R., 2004, New Astr. Rev., 48, 595 (2004)

\bibitem{mattila07} Mattila S., et al. 2007, ApJ, 659, L9 

\bibitem{mattila10} Mattila S., et al. 2010, CBET, 2145 


\bibitem{newton10} Newton J., Puckett T., Orff, T.,  2010, CBET, 2144 

\bibitem{miguel07} P\'erez-Torres, M.A., et al., 2007, ApJ, 671, L21

\bibitem{miguel09} P\'erez-Torres, M.A., et al., 2009, MNRAS, 399. 1641 


\bibitem{ryder10} Ryder S., et al. 2010, CBET, 2189 


\bibitem{thomasson} Thomasson M., et al., 1989, A\&A, 211, 25 

\bibitem{ulvestad} Ulvestad J.S., 2009, AJ, 138, 1529 

\bibitem{petri08a} V\"ais\"anen P., et al. 2008a, MNRAS, 384, 886 

\bibitem{petri08b} V\"ais\"anen P., et al. 2008b, ApJ, 689, L37 

\bibitem{petri09} V\"ais\"anen P., Mattila S., Ryder S., 2010, in: B. Smith et al. (eds), Galaxy Wars, ASPC, 423, 323 (arXiv:0908.3495) 


\end{thebibliography}
\end{document}